\documentclass[aps,prl,showpacs,reprint]{revtex4-1}
\usepackage{amsmath}
\usepackage{amssymb}
\usepackage{graphicx}
\usepackage[colorlinks=true,urlcolor=blue,menucolor=blue,citecolor=black,linkcolor=blue]{hyperref}

\renewcommand{\vec}[1]{\mathbf{#1}}
\newcommand\trick[1]{}

\begin{document}

\title{Superconductivity Induced by Interfacial Coupling to Magnons}
\author{Niklas Rohling, Eirik L\o{}haugen Fj\ae{}rbu, and Arne Brataas}
\affiliation{Department of Physics, Norwegian University of Science and Technology,
NO-7491 Trondheim }

\begin{abstract}
We consider a thin normal metal sandwiched between two ferromagnetic insulators. 
At the interfaces, the exchange coupling causes electrons within the metal to interact with magnons in the insulators. 
This electron-magnon interaction induces electron-electron interactions, which, in turn, can result in p-wave superconductivity. 
In the weak-coupling limit, we solve the gap equation numerically and estimate the critical temperature. 
In YIG-Au-YIG trilayers, superconductivity sets in at temperatures somewhere in the interval between 1 and 10 K. 
EuO-Au-EuO trilayers require a lower temperature, in the range from 0.01 to 1 K. 
\end{abstract}
\maketitle

The interactions between electrons in a conductor and ordered spins across interfaces are of central importance in spintronics \cite{BrunoPRB1995, Tserkovnyak_et_alPRL2002}. 
Here, we focus on the case in which the magnetically ordered system is a ferromagnetic insulator (FI). 
The interaction at an FI-normal metal (NM) interface can be described in terms of an exchange coupling \cite{Tokuyasu_et_alPRB1988,RezendePRB2013,kajiwaraNat2010,TserkovnyakRMP2005}. 
In the static regime, this coupling induces effective Zeeman fields near the boundary 
\cite{Tkaczyk_thesis,Roesler_et_al_1994,Hao_et_al_PRL1991,Miao_et_al_NatComm2013}. 
The magnetization dynamics caused by the coupling can be described in terms of the spin-mixing conductance \cite{RezendePRB2013,kajiwaraNat2010,TserkovnyakRMP2005}. 
Such dynamics can include spin pumping from the FI into the NM \cite{andoApp2010, jungfleischPRB2015} and its reciprocal effect, spin-transfer torques \cite{kajiwaraNat2010, cornelissenNat2015}. 
These spin-transfer torques enable electrical control of the magnetization in FIs \cite{avciNat2016}. 

One important characteristic of FIs is that the Gilbert damping is typically small. 
This leads to low-dissipation magnetization dynamics \cite{sergaApp2010}, which, in turn, facilitates coherent magnon dynamics and the long-range transport of spin signals \cite{kajiwaraNat2010, cornelissenNat2015}. 
These phenomena should also enable other uses of the quantum nature of the magnons.

%
Here, we study a previously unexplored effect that is also governed by the electron-magnon interactions at FI-NM interfaces 
but is qualitatively different from spin pumping and spin-transfer torques. 
We explore how the magnons in FIs can mediate superconductivity in a metal. 
The exchange coupling at the interfaces between the FIs and the NM induces Cooper pairing. 
In this scenario, the electrons and the magnons mediating the pairing reside in two different materials. 
This opens up a wide range of possibilities for tuning the superconducting properties of the system by combining layers with the desired characteristics. 
The electron and magnon dispersions within the layers as well as the electron-magnon coupling between the layers influence the pairing mechanism. 
Consequently, the superconducting gap can also be tuned by modifying the layer thickness, interface quality, and external fields. 

Since the interactions occur at the interfaces, the consequences of the coupling are most profound when the NM layer is thin. 
We therefore consider atomically thin FI and NM layers. 
This also reduces the complexity of the calculations. 
For thicker layers, multiple modes exist along the direction transverse to the interface ($x$), with different effective coupling strengths. 
We expect a qualitatively similar, but somewhat weaker, effect for thicker layers. 

Paramagnonic \cite{KirkpatrickPRB2003} or magnonic \cite{KarchevPRB2003} coupling
may explain experimental observations of superconductivity coexisting with ferromagnetism in bulk materials \cite{SaxenaNat2000, aokiNat2001, pfleidererNat2001}.
Paramagnons \cite{FayPRB1980,KirkpatrickPRB2003} and magnons \cite{KarchevPRB2003,KarchevEPL2015}
are predicted to mediate triplet p-wave pairing with equal and antiparallel spins, respectively.

High-quality thin films offer new possibilities for superconductivity \cite{saitoNat2016}.
Consequently, the emergence of superconductivity at interfaces has recently received considerable attention \cite{Gariglioetal2015, saitoNat2016,ReyrenetalScience2007,Wangetal2012,Boschkeretal2015, Kliminetal2014, Lieetal2014,GongCPL2015, StephanosetalPRB2011}.
Theoretical studies have been conducted on interface-induced superconductivity mediated by phonons \cite{Boschkeretal2015, Kliminetal2014, Lieetal2014}, excitons \cite{Allenderetal1973}, and polarizable localized excitations \cite{KoertingetalPRB2005, StephanosetalPRB2011}. 

A model of interface-induced magnon-mediated d-wave pairing has been proposed to explain the observed superconductivity in Bi/Ni bilayers \cite{GongeSCIENCE2017}. 
A p-wave pairing of electrons with equal momentum---so-called Amperean pairing---has been predicted to occur in a similar system \cite{KagarianPRL2016}. 
Importantly, the electrons that form pairs in these models reside in a spin-momentum-locked surface conduction band.

\begin{figure}
	\def\svgwidth{0.8\columnwidth}
	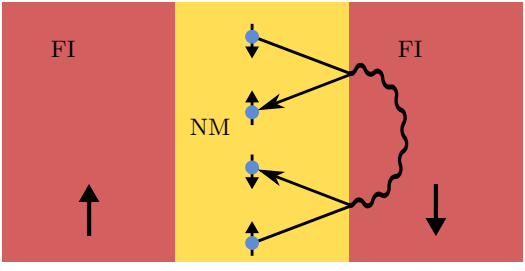
	\caption{A trilayer formed of a normal metal between two ferromagnetic insulators. 
The magnetizations are antiparallel. 
At the interfaces, conduction electrons couple to magnons. 
This results in effective electron-electron interactions in the metal.}
	\label{fig:system}
\end{figure}

By contrast, we consider a spin-degenerate conduction band in an FI-NM-FI trilayer system. 
We find interfacially mediated p-wave superconductivity with antiparallel spins and momenta. 
These pairing symmetries are distinct from those of the 2D systems mentioned above. 
We assume that the equilibrium magnetization of the left (right) FI is along the $\hat{z}$ ($-\hat{z}$) direction; see Fig.~\ref{fig:system}. 
%
We consider matching square lattices, with lattice constant $a$, in all three monolayers. 
The interfacial plane comprises $N$ sites with periodic boundary conditions. The Hamiltonian is 
\begin{equation}
 H = H_{\rm FI}^A + H_{FI}^B + H_{\rm NM} + H_{\rm int} \, ,
\end{equation}
where we use $A$ ($B$) to denote the left (right) FI.  

The Heisenberg Hamiltonian 
\begin{equation}
\label{eq:FI}
 H_{\rm FI}^{A} = - \frac{J}{\hbar^2} \sum_{i}\sum_{j \in {\rm NN}(i)} \vec{S}_i^{A} \cdot \vec S_j^{A} 
\end{equation}
describes the left FI. 
Here, $i$ is an in-plane site, ${\rm NN}(i)$ is the set of its nearest neighbors,
$J$ is the exchange interaction, and $\vec{S}_i^A$ is the localized spin at site $i$. The expression for $H_{\rm FI}^{B}$ is similar. 

For the time being, we assume that the conduction electron eigenstates in the NM are plane waves of the form $c_{\vec{q},\sigma} = \sum_{j} \exp( i\vec r_j\cdot\vec q ) c_{j\sigma}/\sqrt{N}$. 
Here, $c_{j \sigma}^{(\dag)}$ annihilates (creates) a conduction electron with spin $\sigma$ at site $j$ in the NM, and $\vec{q}$ is the wavevector. 
For now, the NM Hamiltonian is $H_{\rm NM}=\sum_{\vec q}\sum_{\sigma}E_{\vec q}c_{\vec q \sigma}^{\dag}c_{\vec q \sigma}$, 
and the dispersion is quadratic, 
\begin{equation}
	E_{\vec q} = \hbar^2 \vec{q}^2 / ( 2m ) \, .  
\end{equation}
Here, $m$ is the effective electron mass. 
Below, when estimating the coupling $J_I$ at YIG-Au interfaces, we consider another Hamiltonian with different eigenstates and a different dispersion. 

We model the coupling between the conduction electrons and the localized spins as an exchange interaction of strength $J_{I}$: 
\begin{equation}
 H_{\rm int} =  - 2 \frac{J_I}{\hbar} \sum_{\sigma \sigma'} \sum_{j} \sum_{L=A,B}
                   c^{\dagger}_{j \sigma}\boldsymbol{\sigma}_{\sigma \sigma'} c_{j \sigma'} \cdot \vec S_j^L \, , 
 \label{eq:int}
\end{equation}
where $\boldsymbol{\sigma} = \left( \sigma_x , \sigma_y, \sigma_z \right)$ is a vector of Pauli matrices. 

After a Holstein-Primakoff transformation, we expand the Heisenberg Hamiltonian given in Eq. \eqref{eq:FI} up to second order in the bosonic operators and diagonalize it.
We represent $\vec{S}_j^A$ by 
$S_{jx}^A+iS_{jy}^A=\hbar\sqrt{2s}a_j$, $S_{jx}^A-iS_{jy}^A=\hbar\sqrt{2s}a_j^\dag$, and $S_{jz}^A=\hbar (s{-}a_j^\dag a_j)$, where $s$ is the spin quantum number of the localized spins 
and $a_j^{(\dag)}$ is a bosonic annihilation (creation) operator at site $j$. 
The magnons in layer $A$, with the form $a_{\vec k} = \sum_{j\in A} \exp( i\vec r_j\cdot\vec k ) a_j/\sqrt{N}$, are the eigenstates of the resulting Hamiltonian. Analogously, the magnons in layer $B$ are denoted by $b_{\vec k}$. 
The magnon dispersion is 
\begin{equation}
	\varepsilon_{\vec k} = 4sJ[2-\cos(k_ya) - \cos(k_za)] \, . 
	\label{eq:magdispersion}
\end{equation}
We disregard second-order terms in the bosonic operators from the interfacial coupling and obtain 
\begin{equation}
	\begin{split}
	H = & \sum_{\vec k} \varepsilon_{\vec k} (a_{\vec k}^\dag a_{\vec k} + b_{\vec k}^\dag b_{\vec k})
    + \sum_{\vec q\sigma}E_{\vec q}c_{\vec q\sigma}^\dag c_{\vec q\sigma}\\
    & + \sum_{\vec k\vec q} V (a_{\vec k}c^{\dag}_{\vec q{+}\vec k,\downarrow}c_{\vec q\uparrow} + b_{\vec k}c^{\dag}_{\vec q{+}\vec k,\uparrow}c_{\vec q\downarrow} ) + \text{h.c.} \, ,
	\end{split}
	\label{eq:magnonHamiltonian}
\end{equation}
where $V= -2J_I\sqrt{s}/\sqrt{2N}$ is the coupling strength between the electrons in the NM and the magnons in the FI layers. 

There is no induced Zeeman field in the NM since the magnetizations in the FIs are antiparallel. 
Analogously to phonon-mediated coupling in conventional superconductors, the magnons mediate effective interactions between the electrons. 
For electron pairs with opposite momenta, we obtain
\begin{equation}
H_{\rm pair} = \sum_{\vec k\vec k'}V_{\vec k \vec k'}
                                   c_{\vec k\downarrow}^\dag
                                   c_{-\vec k\uparrow}^\dag
                                   c_{-\vec k'\uparrow}
                                   c_{\vec k'\downarrow} \, ,
\end{equation}
with the interaction strength 
\begin{equation}
V_{\vec k \vec k'} = 2|V|^2
                      \frac{\varepsilon_{\vec k{+}\vec k'}}
                           {\varepsilon_{\vec k{+}\vec k'}^2 - (E_{\vec k} - E_{\vec k'})^2} \, .
\label{eq:intstrength}
\end{equation}
We define the gap function in the usual way: $\Delta_{\vec k} = \sum_{\vec k'} V_{\vec k \vec k'}\langle c_{-\vec k'\uparrow}c_{\vec k'\downarrow}\rangle$. The gap equation becomes 
\begin{equation}
 \Delta_{\vec k} = - \sum_{\vec k'} V_{\vec k \vec k'} \frac{\Delta_{\vec k'}}{2\tilde E_{\vec k'}}
                                    \tanh\left(\frac{\tilde E_{\vec k'}}{2k_B T}\right) \, , 
\label{eq:gap}
\end{equation}
where $\tilde E_{\vec k} = \sqrt{(E_{\vec k}-E_F)^2+|\Delta_{\vec k}|^2}$ and $E_F$ is the Fermi energy. 

In the continuum limit, we replace the discrete sum over momenta $\vec{k}$ with integrals over $E=E_{\vec{k}}$ and the angle $\varphi$, where $\vec{k} = k \left[\sin(\varphi),\cos(\varphi) \right]$. 
We assume that only the conduction electrons close to the Fermi surface form pairs. 
The magnon energy that appears in Eq.\ (\ref{eq:intstrength}) is then given by $\varepsilon_{\vec k{+}\vec k'}\approx \varepsilon(\varphi',\varphi)$, where 
\begin{equation}
	\begin{split}
		\varepsilon(\varphi',\varphi) =& 4sJ \{ 2 - \cos(k_F a [\sin\varphi{+}\sin\varphi']) \\
		& -\cos(k_Fa[\cos\varphi{+}\cos\varphi']) \} \, . 
		\end{split}
		\label{eq:magnonfull}
\end{equation}
Here, $k_F=\sqrt{2mE_F}/\hbar$ is the Fermi wavenumber. We assume that the NM is half filled, $k_F=\sqrt{2\pi}/a$. 
We introduce the energy scale $E^*=4 s J k_F^2 a^2=8\pi sJ$, which is associated with the FI exchange interaction. 
Then, we scale all other energies with respect to $E^*$: 
$\delta = \Delta/E^*$, $\tau = k_BT/E^*$, $x = (E-E_F)/E^*$, $\tilde x = \tilde E/E^*$, and $\epsilon = \varepsilon/E^*$. 
In this way, the gap equation presented in Eq. \eqref{eq:gap} simplifies to 
\begin{equation}
	\label{eq:gap_int}
 \delta(x,\varphi) =  \frac{-\sqrt{2}\alpha}{\pi}
                     \!\!\int\limits_{-x_B}^{x_B}\!\!\!\!dx'
                      \!\!\int\limits_0^{2\pi}\!\!d\varphi'
                      \frac{\epsilon(\varphi'\!,\varphi)\delta(x'\!,\varphi')
                                 \tanh\!\!\left[\frac{\tilde x'}{2\tau}\right]}
                           {\tilde x'[\epsilon^2(\varphi'\!,\varphi)-(x{-}x')^2]} \, ,
\end{equation}
with the dimensionless coupling constant $\alpha = J_I^2/(16\sqrt{2}\pi E_F J) = J_I^2ma^2/(16\sqrt{2}\pi^2\hbar^2J)$. 
In Eq.~\eqref{eq:gap_int}, we have restricted the energy integral to the range $[ E_F-x_B E^*, E_F+x_B E^* ]$. 
We choose $x_B$---based on the value of $\alpha$---in the following way. 
$x_B$ must be sufficiently large that all contributions to the gap from regions outside this range are vanishingly small. 
In the weak-coupling limit ($\alpha \ll 1$), the gap function has a narrow peak near $x = 0$, and therefore, $x_{B}$ can be much smaller than $1$. 

To gain a better understanding, we first assume a quadratic dispersion for the magnons, which matches that of Eq.\ (\ref{eq:magdispersion}) in the long-wavelength limit. 
Consequently, the dimensionless magnon energy $\epsilon(\varphi',\varphi)$ becomes $\epsilon_q(\varphi',\varphi) = 1+\cos(\varphi'{-}\varphi)$. 
Below, we numerically check the correspondence between the solutions resulting from the full dispersion versus the solutions obtained with the quadratic approximation assumed here. 
For the quadratic magnon dispersion, the gap equation has a solution with p-wave symmetry, $\delta(x,\varphi) = f(x)\exp(\pm i\varphi)$. 
Applying this ansatz to Eq.~(\ref{eq:gap_int}), we calculate the integral over the angle $\varphi'$ in the weak-coupling limit 
\footnote{ 
For the integration, we use the Cauchy principle value and the fact that $\Delta x=|x-x'|\ll1$,
\begin{equation*}
\begin{split}
V(\Delta x) = &  - \frac{\sqrt{2}}{\pi}\int_0^{2\pi}d\tilde\varphi\,
                         \frac{(1+\cos\tilde\varphi)\cos\tilde\varphi}
                              {(1+\cos\tilde\varphi)^2-\Delta x^2}\\
          = & \frac{1+|\Delta x|}{\sqrt{\Delta x^2/2+\Delta x}} - 2\sqrt{2} \approx \frac{1}{\sqrt{\Delta x}}-2\sqrt{2} \, .
\end{split}
\end{equation*} \protect\trick.
}. 
The gap equation becomes 
\begin{equation}
\label{eq:gap_simp}
 f(x) = \alpha \int\limits_{-x_B}^{x_B}\!\!dx'\, V(x{-}x')
                                  \frac{f(x')\tanh\left[\frac{\sqrt{x'^2+f(x')^2}}{2\tau}\right]}
                                       {\sqrt{x'^2+f(x')^2}} \, ,
\end{equation}
where $V(x{-}x') \approx 1/\sqrt{|x{-}x'|} - 2\sqrt{2} $. 
%

\begin{figure}
\includegraphics[width=\columnwidth]{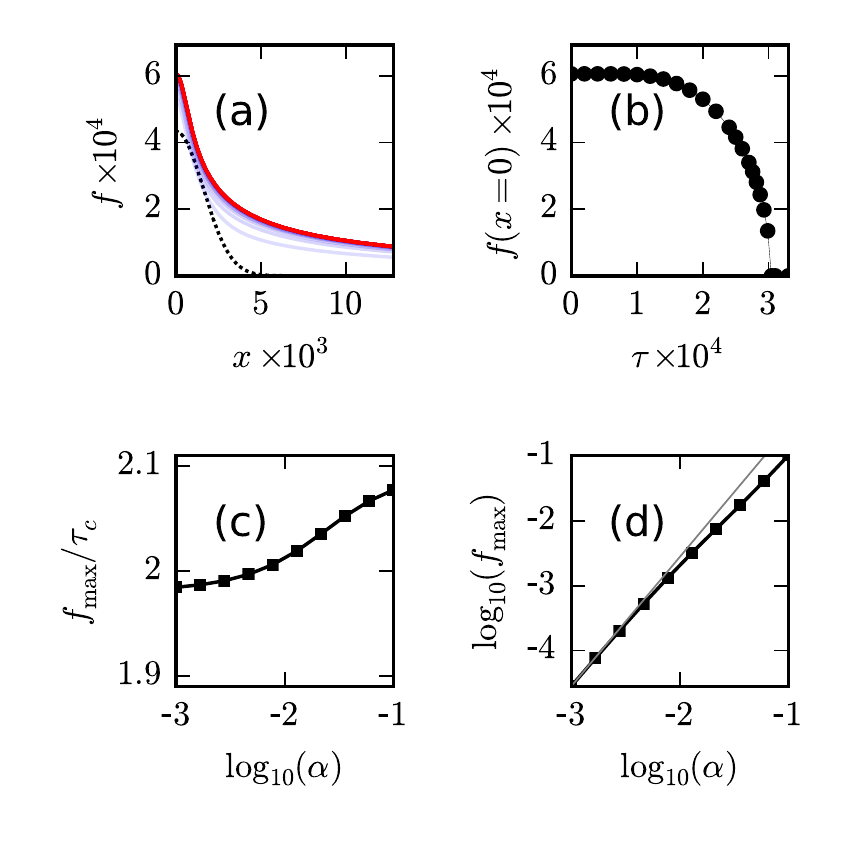}
\caption{Numerical solutions to the gap equation (\ref{eq:gap_simp}) determined through iteration. 
(a) Gaussian-shaped initial guess (dashed line) and the results of the first eight iterative calculations of the gap $f(x)$ (from light blue to red) when the dimensionless temperature is $\tau=0$ and the coupling constant is $\alpha = 0.005$. 
Note that $f(-x)=f(x)$ and that the energy cutoff $x_B \approx 0.03$ lies outside the range of the plot. 
(b) Gap $f$ at energy $x=0$ as a function of $\tau$ for $\alpha=0.005$. 
(c) Ratio between the maximum gap value, $f_{\rm{max}}$, and the dimensionless critical temperature $\tau_c$ as a function of $\alpha$. 
(d) $\alpha$ dependence of $f_{\rm max}$. 
The gray line corresponds to a quadratic dependence, $f_{\rm max}\sim\alpha^2$. 
}
\label{fig:gap_temp}
\end{figure}

Using a Gaussian centered at $x=0$ as an initial guess, we solve Eq.~(\ref{eq:gap_simp}) numerically through iteration 
\footnote{
To eliminate the singularity in $V(x{-}x')$ at $x'=x$ for the numerical integration, 
we replace $V(x{-}x')$ in Eq.~(\ref{eq:gap_simp}) with $\int_0^x d\tilde{x} V(\tilde{x}{-}x')$ and $f(x)$ on the left-hand side of Eq.~(\ref{eq:gap_simp}) with $F(x) = \int_0^x d\tilde{x} f(\tilde{x})$. 
In each iteration, we numerically evaluate the integral over $x'$ in the resulting equation
and obtain $f(x)$ by numerically differentiating $F$. 
}.
Fig.~\ref{fig:gap_temp} shows the results. 
For a fixed coupling $\alpha$, the maximum value occurs when $x=0$ and $\tau=0$. 
The dimensionless critical temperature $\tau_c$ is the temperature at which the gap vanishes. 
As in the BCS theory, the gap equation can also be solved analytically by approximating $V(x)$ as a constant with a cutoff centered at $x = 0$. 
In this constant-potential approximation, the ratio $f_{\rm max}/\tau_c$ is approximately $1.76$, which is slightly lower than what we find numerically; see Fig.~\ref{fig:gap_temp} (c). 

\begin{figure}
\includegraphics[width=\columnwidth]{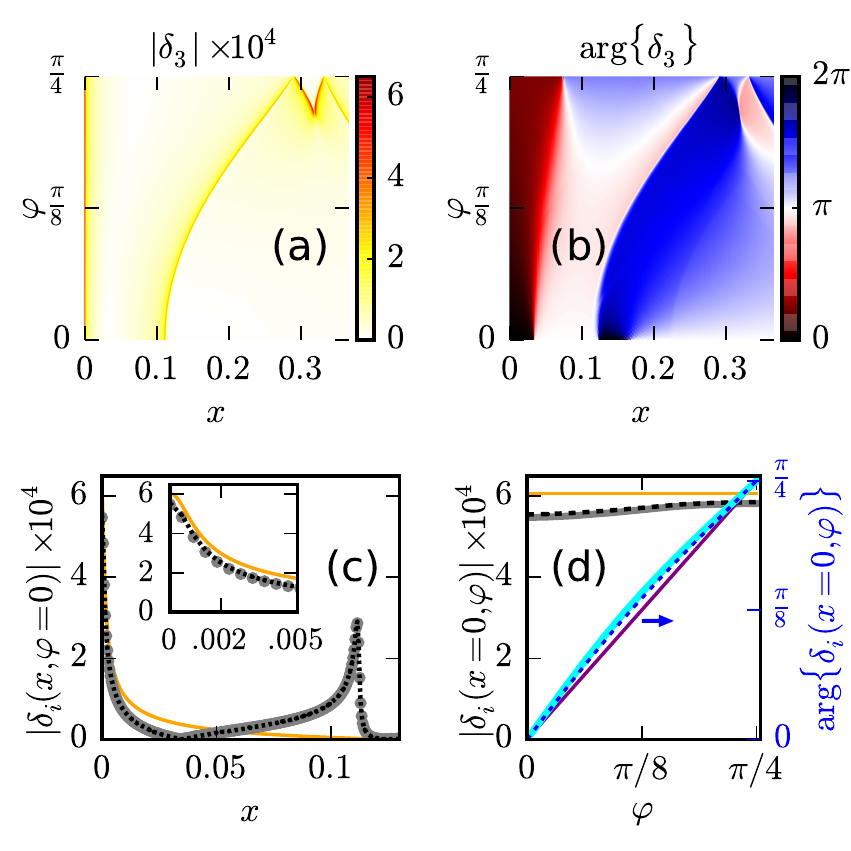}
\caption{Numerical iteration of the gap equation (\ref{eq:gap_int}), starting from the 
solution to Eq.~(\ref{eq:gap_simp}) as an initial guess,
for $\tau=0$ and $\alpha=0.005$. 
(a,b) Absolute value and phase of $\delta_3(x,\varphi)$, where the index $3$ indicates the number of iterations. 
(c) $|\delta_i(x,\varphi=0)|$ for $i=0$ (orange line), $i=2$ (black dashed line), and $i=3$ (gray circles). 
(d) $|\delta_i(x=0,\varphi)|$ (left axis) for $i=0,2,3$, with the same colors as in (c), and
the phase of $\delta_i(x=0,\varphi)$ (right axis) for $i=0$ (purple), $i=2$ (blue, dashed), $i=3$ (cyan, wide). 
Note that the difference from the second to the third iteration is nearly indiscernible. 
}
\label{fig:num_check}
\end{figure}

Let us check that the numerical solutions to Eq.~(\ref{eq:gap_simp}), for the quadratic magnon energy, 
resemble the solutions to Eq.~(\ref{eq:gap_int}) for the full magnon energy of Eq.\ (\ref{eq:magnonfull}). 
To this end, we numerically iterate Eq.~(\ref{eq:gap_int}), starting from the solution to Eq.~(\ref{eq:gap_simp}) as the initial guess 
\footnote{In the same way as for the iteration of Eq.~(\ref{eq:gap_simp}), 
we eliminate singularities from the integral over $x'$ in Eq.~(\ref{eq:gap_int}). 
Hence, we replace the factor $U(\varphi',\varphi,x',x) = 1/[\epsilon^2(\varphi',\varphi)-(x{-}x')^2]$ in the integrand with $\int_{0}^{x} d\tilde{X} \int_{0}^{\tilde{X}} d\tilde{x} U(\varphi',\varphi,x',\tilde{x})$ and $\delta(x,\varphi)$ with $D(x,\varphi) = \int_{0}^{x} d\tilde{X} \int_{0}^{\tilde{X}} d\tilde{x} \delta(\tilde{x},\varphi)$. 
In each iteration, 
we find $D$ by numerically integrating over $x'$ and then find $\delta$ by numerically differentiating $D$ twice.
}. 
We consider the case of zero temperature, $\tau=0$. 
The symmetries $\delta(x,\varphi)=\delta(-x,\varphi)=i\delta(x,\varphi+\pi/2)=\delta^*(x,-\varphi)$, 
where $\delta^*$ is the complex conjugate of $\delta$, imply that we need to consider only $x>0$ and $0<\varphi<\pi/4$. 
We show the results of these iterative calculations in Fig.~\ref{fig:num_check}. 
The third iteration of $\delta$ is shown in Fig.~\ref{fig:num_check} (a,b). 
After only three iterations, the differences between consecutive functions are already nearly imperceptible; see Fig.~\ref{fig:num_check} (c,d). 
The gap as a function of energy still exhibits a peak at the Fermi energy.
Compared with the results obtained for a quadratic magnon dispersion, this peak is of a similar shape but is slightly lower and narrower; see the inset of Fig.~\ref{fig:num_check} (c). 
There are also additional features of $\delta(x,\varphi)$ at positions $(x,\varphi) = (\epsilon(\varphi',\varphi),\varphi)$ 
in the parameter space where the derivative of $\epsilon(\varphi',\varphi)$ with respect to $\varphi'$ vanishes. 

Next, we estimate the critical temperatures $T_c$ for two possible experimental realizations, one in which the FI is yttrium-iron-garnet (YIG) and one in which the FI is europium oxide (EuO). 
The NM layer is gold in both cases. 
We consider the YIG-Au-YIG trilayer first. 

For the FIs, we assume---encouraged by the results presented in Fig.~\ref{fig:num_check}---that the low-energy magnons dominate the gap. 
The relevant magnons can therefore be well described by a quadratic dispersion. 
Our model assumes that the FI and NM layers have the same lattice structure.
However, in reality, the unit cell of YIG is much larger than that of Au. 
To capture the properties of YIG in our model, 
we fit the parameters such that the FIs have the same exchange stiffness ($D / k_B = 71$~$\rm{K}\,\rm{nm}^2$ \cite{klinglerApp2015}) and saturation magnetization ($M_s= 1.6 \cdot 10^5$~A/m \cite{klinglerApp2015}) as those of bulk YIG. 
We assume that each YIG layer has a thickness equal to the bulk lattice constant of YIG ($ a_{\rm YIG} \approx 12$~\r{A} \cite{klinglerApp2015}). 
We use the thickness, the saturation magnetization and the electron gyromagnetic ratio $\gamma_e$ to estimate the spin quantum number $s=M_s a_{\rm YIG}a^2/(\hbar\gamma_e)$. 
Using the quadratic dispersion approximation, we determine the exchange interaction to be $J=D/(2a^2s)$. 
The lattice spacing $a$ remains undetermined. 

In the bulk, gold has an fcc lattice and a half-filled conduction band. 
We use experimental values of the Fermi energy ($E^{\rm{B}}_{F} = 5.5$~eV \cite{ashcroft1976}) and the Sharvin conductance ($g_{\rm{Sh}} = 12$~$\rm{nm}^{-2}$ \cite{TserkovnyakRMP2005}) to determine the effective mass, $m=2\pi g_{\rm Sh}\hbar^2/E^{\rm{B}}_F$. 
We assume that the monolayer is half filled and has the same effective electron mass as that of bulk gold. 
We consider the case in which the monolayer lattice constant $a$ is equal to the lattice constant $a_{\rm{t}}$ of a simple cubic tight-binding model for gold. 
$a_{\rm{t}}$ is approximately $20 \%$ smaller than the bulk nearest-neighbor distance of actual gold. 

We calculate the interfacial exchange coupling $J_I$ for a YIG-Au bilayer in terms of the spin-mixing conductance, which has been experimentally measured. 
In doing this calculation, we use the same model for the YIG as in the trilayer case; however, for the gold, we employ a tight-binding model of the form
$H_{\rm{t}} = - t_{\rm{t}} \sum_{\sigma} \sum_{i} \sum_{j \in {\rm NN}(i)} c^{\dagger}_{i \sigma} c_{j \sigma}$, with a simple cubic lattice. 
The Hamiltonian of the bilayer is $H_{\rm B} = H_{\rm{t}} + H_{\rm FI}^A + H_{\rm int}$.
We assume that $J_I s \ll t_{\rm{t}}$, which allows us to disregard the proximity-induced Zeeman field. 
The energy eigenstates $c^{\rm{t}}_{\vec{q} \sigma}$ and the dispersion $E^{\rm{t}}_{\vec{q}} = 4 t_{\rm{t}} \left( 3 - \cos(q_x a_{\rm{t}}) - \cos(q_y a_{\rm{t}}) - \cos(q_z a_{\rm{t}}) \right)$ of $H_{\rm{t}}$ are well known. 
Under the assumption of half filling, we find that $ t_{\rm{t}} = E^{\rm{B}}_{F}/12 $ and $a_{\rm{t}} = \sqrt{0.63/g_{\rm{Sh}} }$. 
We use the same experimental values for $E^{\rm{B}}_{F}$ and $g_{\rm{Sh}}$ (from Ref. \onlinecite{ashcroft1976,TserkovnyakRMP2005}) as before. 

We set the lattice constant of the trilayer, $a$, equal to the lattice constant of the bilayer, $a_{\rm{t}}$. 
This ensures that both models have the same lattice structure at the interface and, consequently, that the interfacial exchange interaction Hamiltonian $H_{\rm{int}}$ has the same form in both cases. 
To first order in the bosonic operators, $H_{\rm{int}}= \sum_{\vec k\vec q} V_{\rm{t}} a_{\vec k}c^{\rm{t} \dag}_{\vec q{+}\vec k,\downarrow}c^{\rm{t}}_{\vec q\uparrow} $. 
The coupling strength $V_{\rm{t}}$ is proportional to the amplitudes of the tight-binding-model eigenstates at the interface: $V_{\rm{t}} = 2 V \sin(q_x a_{\rm{t}}) \sin([k_x + q_x]a_{\rm{t}})$.
The spin-mixing conductance can now be calculated for the ferromagnetic resonance (FMR) mode, resulting \cite{benderPRL2012} in $g_{\uparrow \downarrow} = 4 a_{t}^2 V_0 s N / \left(2 \pi\right)^2$, where 
\begin{equation}
	\begin{split}
		V_0 &= \iint \! \! |V|^2 \sin(q_x a_{\rm{t}})^2 \sin(q_x' a_{\rm{t}})^2 \delta \left( q_y - q_y' \right) \\
		&  \delta \left( q_z - q_z' \right) \delta\left(E^{\rm{t}}_{\vec{q}} - E_{F} \right) \delta\left(E^{\rm{t}}_{\vec{q}'} - E_{F} \right) \text{d}^{3}\vec{q}\text{d}^{3}\vec{q}' \, .
	\end{split}
	\label{eq:spinmixintegral}
\end{equation}
We numerically evaluate $V_0$ and estimate the bilayer interfacial exchange coupling $J_{I} = \sqrt{(2 \pi)^2 g_{\uparrow \downarrow} t_{\rm{t}}^2 a_{\rm{t}}^2/(9.16 s^2) }$ using measured values of the spin-mixing conductance $g_{\uparrow \downarrow}$. 
We assume that $J_I$ has the same value in the trilayer case. 
Using $E^*=8 \pi s J$, we find that $E^*$ is approximately $1.5$~eV. 
We find the coupling constant $\alpha$ from the relation $\alpha = J_I^2ma^2/(16\sqrt{2}\pi^2\hbar^2J)$. 
The reported experimental values for the spin-mixing conductance range from $1.2$~$\rm{nm}^{-2}$ to $6$~$\rm{nm}^{-2}$ \cite{heinrichPRL2011,BurrowesApp2012,haertingerPRB2015}. 
In turn, this implies that $\alpha$ lies in the range of $[0.0014\text{--}0.007]$.  
The corresponding critical temperatures range from $0.5$~K to $10$~K. 

Next, we consider a EuO-Au-EuO trilayer. 
Europium oxide has an fcc lattice structure with a lattice constant of $5.1$ \r{A}, a spin quantum number of $s=7/2$ and a nearest-neighbor exchange coupling of $J/k_B = 0.6$~K \cite{Mauger_Godart_1986}. 
The nodes on a $(100)$ surface of an fcc lattice form a square lattice in which the lattice constant is equal to the distance between nearest neighbors in the bulk. 
We assume that the monolayer has the same structure 
and therefore set $a$ equal to the distance between nearest neighbors in bulk EuO. 
We use the same effective mass as for the YIG-Au-YIG trilayer. 
Then, the Fermi energy is $E_F = 1.8$~eV, and the energy scale $E^*/k_B$ is approximately $53$~K. 
Values on the order of $10$ meV have been reported for the interfacial exchange coupling strengths $J_I$ 
\footnote{
Note that the strength of the exchange coupling is defined differently in 
Refs.~\cite{Tkaczyk_thesis,Roesler_et_al_1994}; the values reported
there must be divided by 2 to obtain the value of $J_I$ as it is
defined in this Letter.
}
in EuO/Al \cite{Tkaczyk_thesis}, EuO/V \cite{Roesler_et_al_1994}, and EuS/Al \cite{Hao_et_al_PRL1991,Miao_et_al_NatComm2013}. 
These estimates were based on measurements of a proximity-induced effective Zeeman field. 
Under the assumption that $J_I$ is in the range of $[5\text{--}15]$~meV, 
we find a wide range of values of $[0.004\text{--}0.03]$ for $\alpha$. 
We estimate the corresponding critical temperatures numerically using the quadratic dispersion approximation. 
Finally, we find a range of $[0.01\text{--}0.4]$~K as possible values for $T_c$. 

In conclusion, interfacial coupling to magnons induces p-wave superconductivity in metals. 
The critical temperatures are experimentally accessible in the weak-coupling limit. 
The gap size strongly depends on the magnitude of the interfacial exchange coupling. 
The thickness dependence, the robustness against disorder, and the physics beyond the weak-coupling limit should be explored in the future. 

This work was partially supported by the European Research Council via Advanced Grant No. 669442 ``Insulatronics'' and the Research Council of Norway via the Centre of Excellence ``QuSpin''.


\begin{thebibliography}{46}%
\makeatletter
\providecommand \@ifxundefined [1]{%
 \@ifx{#1\undefined}
}%
\providecommand \@ifnum [1]{%
 \ifnum #1\expandafter \@firstoftwo
 \else \expandafter \@secondoftwo
 \fi
}%
\providecommand \@ifx [1]{%
 \ifx #1\expandafter \@firstoftwo
 \else \expandafter \@secondoftwo
 \fi
}%
\providecommand \natexlab [1]{#1}%
\providecommand \enquote  [1]{``#1''}%
\providecommand \bibnamefont  [1]{#1}%
\providecommand \bibfnamefont [1]{#1}%
\providecommand \citenamefont [1]{#1}%
\providecommand \href@noop [0]{\@secondoftwo}%
\providecommand \href [0]{\begingroup \@sanitize@url \@href}%
\providecommand \@href[1]{\@@startlink{#1}\@@href}%
\providecommand \@@href[1]{\endgroup#1\@@endlink}%
\providecommand \@sanitize@url [0]{\catcode `\\12\catcode `\$12\catcode
  `\&12\catcode `\#12\catcode `\^12\catcode `\_12\catcode `\%12\relax}%
\providecommand \@@startlink[1]{}%
\providecommand \@@endlink[0]{}%
\providecommand \url  [0]{\begingroup\@sanitize@url \@url }%
\providecommand \@url [1]{\endgroup\@href {#1}{\urlprefix }}%
\providecommand \urlprefix  [0]{URL }%
\providecommand \Eprint [0]{\href }%
\providecommand \doibase [0]{http://dx.doi.org/}%
\providecommand \selectlanguage [0]{\@gobble}%
\providecommand \bibinfo  [0]{\@secondoftwo}%
\providecommand \bibfield  [0]{\@secondoftwo}%
\providecommand \translation [1]{[#1]}%
\providecommand \BibitemOpen [0]{}%
\providecommand \bibitemStop [0]{}%
\providecommand \bibitemNoStop [0]{.\EOS\space}%
\providecommand \EOS [0]{\spacefactor3000\relax}%
\providecommand \BibitemShut  [1]{\csname bibitem#1\endcsname}%
\let\auto@bib@innerbib\@empty
\bibitem [{\citenamefont {Bruno}(1995)}]{BrunoPRB1995}%
  \BibitemOpen
  \bibfield  {author} {\bibinfo {author} {\bibfnamefont {P.}~\bibnamefont
  {Bruno}},\ }\href {\doibase 10.1103/PhysRevB.52.411} {\bibfield  {journal}
  {\bibinfo  {journal} {Phys. Rev. B}\ }\textbf {\bibinfo {volume} {52}},\
  \bibinfo {pages} {411} (\bibinfo {year} {1995})}\BibitemShut {NoStop}%
\bibitem [{\citenamefont {Tserkovnyak}\ \emph {et~al.}(2002)\citenamefont
  {Tserkovnyak}, \citenamefont {Brataas},\ and\ \citenamefont
  {Bauer}}]{Tserkovnyak_et_alPRL2002}%
  \BibitemOpen
  \bibfield  {author} {\bibinfo {author} {\bibfnamefont {Y.}~\bibnamefont
  {Tserkovnyak}}, \bibinfo {author} {\bibfnamefont {A.}~\bibnamefont
  {Brataas}}, \ and\ \bibinfo {author} {\bibfnamefont {G.~E.~W.}\ \bibnamefont
  {Bauer}},\ }\href {\doibase 10.1103/PhysRevLett.88.117601} {\bibfield
  {journal} {\bibinfo  {journal} {Phys. Rev. Lett.}\ }\textbf {\bibinfo
  {volume} {88}},\ \bibinfo {pages} {117601} (\bibinfo {year}
  {2002})}\BibitemShut {NoStop}%
\bibitem [{\citenamefont {Tokuyasu}\ \emph {et~al.}(1988)\citenamefont
  {Tokuyasu}, \citenamefont {Sauls},\ and\ \citenamefont
  {Rainer}}]{Tokuyasu_et_alPRB1988}%
  \BibitemOpen
  \bibfield  {author} {\bibinfo {author} {\bibfnamefont {T.}~\bibnamefont
  {Tokuyasu}}, \bibinfo {author} {\bibfnamefont {J.~A.}\ \bibnamefont {Sauls}},
  \ and\ \bibinfo {author} {\bibfnamefont {D.}~\bibnamefont {Rainer}},\ }\href
  {\doibase 10.1103/PhysRevB.38.8823} {\bibfield  {journal} {\bibinfo
  {journal} {Phys. Rev. B}\ }\textbf {\bibinfo {volume} {38}},\ \bibinfo
  {pages} {8823} (\bibinfo {year} {1988})}\BibitemShut {NoStop}%
\bibitem [{\citenamefont {Rezende}\ \emph {et~al.}(2013)\citenamefont
  {Rezende}, \citenamefont {Rodr{\'{\i}}guez-Su{\'a}rez},\ and\ \citenamefont
  {Azevedo}}]{RezendePRB2013}%
  \BibitemOpen
  \bibfield  {author} {\bibinfo {author} {\bibfnamefont {S.~M.}\ \bibnamefont
  {Rezende}}, \bibinfo {author} {\bibfnamefont {R.~L.}\ \bibnamefont
  {Rodr{\'{\i}}guez-Su{\'a}rez}}, \ and\ \bibinfo {author} {\bibfnamefont
  {A.}~\bibnamefont {Azevedo}},\ }\href
  {https://doi.org/10.1103/PhysRevB.88.014404} {\bibfield  {journal} {\bibinfo
  {journal} {Phys. Rev. B}\ }\textbf {\bibinfo {volume} {88}},\ \bibinfo
  {pages} {014404} (\bibinfo {year} {2013})}\BibitemShut {NoStop}%
\bibitem [{\citenamefont {Kajiwara}\ \emph {et~al.}(2010)\citenamefont
  {Kajiwara}, \citenamefont {Harii}, \citenamefont {Takahashi}, \citenamefont
  {Ohe}, \citenamefont {Uchida}, \citenamefont {Mizuguchi}, \citenamefont
  {Umezawa}, \citenamefont {Kawai}, \citenamefont {Ando}, \citenamefont
  {Takanash}, \citenamefont {Maekawa},\ and\ \citenamefont
  {Saitoh}}]{kajiwaraNat2010}%
  \BibitemOpen
  \bibfield  {author} {\bibinfo {author} {\bibfnamefont {Y.}~\bibnamefont
  {Kajiwara}}, \bibinfo {author} {\bibfnamefont {K.}~\bibnamefont {Harii}},
  \bibinfo {author} {\bibfnamefont {S.}~\bibnamefont {Takahashi}}, \bibinfo
  {author} {\bibfnamefont {J.}~\bibnamefont {Ohe}}, \bibinfo {author}
  {\bibfnamefont {K.}~\bibnamefont {Uchida}}, \bibinfo {author} {\bibfnamefont
  {M.}~\bibnamefont {Mizuguchi}}, \bibinfo {author} {\bibfnamefont
  {H.}~\bibnamefont {Umezawa}}, \bibinfo {author} {\bibfnamefont
  {H.}~\bibnamefont {Kawai}}, \bibinfo {author} {\bibfnamefont
  {K.}~\bibnamefont {Ando}}, \bibinfo {author} {\bibfnamefont {K.}~\bibnamefont
  {Takanash}}, \bibinfo {author} {\bibfnamefont {S.}~\bibnamefont {Maekawa}}, \
  and\ \bibinfo {author} {\bibfnamefont {E.}~\bibnamefont {Saitoh}},\ }\href
  {https://dx.doi.org/10.1038/nature08876} {\bibfield  {journal} {\bibinfo
  {journal} {Nature}\ }\textbf {\bibinfo {volume} {464}},\ \bibinfo {pages}
  {7286} (\bibinfo {year} {2010})}\BibitemShut {NoStop}%
\bibitem [{\citenamefont {Tserkovnyak}\ \emph {et~al.}(2005)\citenamefont
  {Tserkovnyak}, \citenamefont {Brataas}, \citenamefont {Bauer},\ and\
  \citenamefont {Halperin}}]{TserkovnyakRMP2005}%
  \BibitemOpen
  \bibfield  {author} {\bibinfo {author} {\bibfnamefont {Y.}~\bibnamefont
  {Tserkovnyak}}, \bibinfo {author} {\bibfnamefont {A.}~\bibnamefont
  {Brataas}}, \bibinfo {author} {\bibfnamefont {G.~E.~W.}\ \bibnamefont
  {Bauer}}, \ and\ \bibinfo {author} {\bibfnamefont {B.~I.}\ \bibnamefont
  {Halperin}},\ }\href {https://doi.org/10.1103/RevModPhys.77.1375} {\bibfield
  {journal} {\bibinfo  {journal} {Rev. Mod. Phys.}\ }\textbf {\bibinfo {volume}
  {77}},\ \bibinfo {pages} {1375} (\bibinfo {year} {2005})}\BibitemShut
  {NoStop}%
\bibitem [{\citenamefont {Tkaczyk}(1988)}]{Tkaczyk_thesis}%
  \BibitemOpen
  \bibfield  {author} {\bibinfo {author} {\bibfnamefont {J.~E.}\ \bibnamefont
  {Tkaczyk}},\ }\href {http://hdl.handle.net/1721.1/17231} {Ph.D. thesis},\
  \bibinfo  {school} {MIT} (\bibinfo {year} {1988})\BibitemShut {NoStop}%
\bibitem [{\citenamefont {Roesler}\ \emph {et~al.}(1994)\citenamefont
  {Roesler}, \citenamefont {Filipkowski}, \citenamefont {Broussard},
  \citenamefont {Idzerda}, \citenamefont {Osofsky},\ and\ \citenamefont
  {Soulen}}]{Roesler_et_al_1994}%
  \BibitemOpen
  \bibfield  {author} {\bibinfo {author} {\bibfnamefont {G.~M.}\ \bibnamefont
  {Roesler}}, \bibinfo {author} {\bibfnamefont {M.~E.}\ \bibnamefont
  {Filipkowski}}, \bibinfo {author} {\bibfnamefont {P.~R.}\ \bibnamefont
  {Broussard}}, \bibinfo {author} {\bibfnamefont {Y.~U.}\ \bibnamefont
  {Idzerda}}, \bibinfo {author} {\bibfnamefont {M.~S.}\ \bibnamefont
  {Osofsky}}, \ and\ \bibinfo {author} {\bibfnamefont {R.~J.}\ \bibnamefont
  {Soulen}},\ }\href {http://dx.doi.org/10.1117/12.179175} {\bibfield
  {journal} {\bibinfo  {journal} {Proc.~SPIE}\ }\textbf {\bibinfo {volume}
  {2157}},\ \bibinfo {pages} {285} (\bibinfo {year} {1994})}\BibitemShut
  {NoStop}%
\bibitem [{\citenamefont {Hao}\ \emph {et~al.}(1991)\citenamefont {Hao},
  \citenamefont {Moodera},\ and\ \citenamefont {Meservey}}]{Hao_et_al_PRL1991}%
  \BibitemOpen
  \bibfield  {author} {\bibinfo {author} {\bibfnamefont {X.}~\bibnamefont
  {Hao}}, \bibinfo {author} {\bibfnamefont {J.~S.}\ \bibnamefont {Moodera}}, \
  and\ \bibinfo {author} {\bibfnamefont {R.}~\bibnamefont {Meservey}},\ }\href
  {https://doi.org/10.1103/PhysRevLett.67.1342} {\bibfield  {journal} {\bibinfo
   {journal} {Phys.~Rev.~Lett.}\ }\textbf {\bibinfo {volume} {67}},\ \bibinfo
  {pages} {1342} (\bibinfo {year} {1991})}\BibitemShut {NoStop}%
\bibitem [{\citenamefont {Miao}\ \emph {et~al.}(2014)\citenamefont {Miao},
  \citenamefont {Chang}, \citenamefont {Assaf}, \citenamefont {Heiman},\ and\
  \citenamefont {Moodera}}]{Miao_et_al_NatComm2013}%
  \BibitemOpen
  \bibfield  {author} {\bibinfo {author} {\bibfnamefont {G.-X.}\ \bibnamefont
  {Miao}}, \bibinfo {author} {\bibfnamefont {J.}~\bibnamefont {Chang}},
  \bibinfo {author} {\bibfnamefont {B.~A.}\ \bibnamefont {Assaf}}, \bibinfo
  {author} {\bibfnamefont {D.}~\bibnamefont {Heiman}}, \ and\ \bibinfo {author}
  {\bibfnamefont {J.~S.}\ \bibnamefont {Moodera}},\ }\href
  {http://dx.doi.org/10.1038/ncomms4682} {\bibfield  {journal} {\bibinfo
  {journal} {Nature Communications}\ }\textbf {\bibinfo {volume} {5}},\
  \bibinfo {pages} {3682} (\bibinfo {year} {2014})}\BibitemShut {NoStop}%
\bibitem [{\citenamefont {Ando}\ \emph {et~al.}(2010)\citenamefont {Ando},
  \citenamefont {Takahashi}, \citenamefont {Ieda}, \citenamefont {Kajiwara},
  \citenamefont {Nakayama}, \citenamefont {Yoshino}, \citenamefont {Harii},
  \citenamefont {Fujikawa}, \citenamefont {Matsuo}, \citenamefont {Maekawa},\
  and\ \citenamefont {Saitoh}}]{andoApp2010}%
  \BibitemOpen
  \bibfield  {author} {\bibinfo {author} {\bibfnamefont {K.}~\bibnamefont
  {Ando}}, \bibinfo {author} {\bibfnamefont {S.}~\bibnamefont {Takahashi}},
  \bibinfo {author} {\bibfnamefont {J.}~\bibnamefont {Ieda}}, \bibinfo {author}
  {\bibfnamefont {Y.}~\bibnamefont {Kajiwara}}, \bibinfo {author}
  {\bibfnamefont {H.}~\bibnamefont {Nakayama}}, \bibinfo {author}
  {\bibfnamefont {T.}~\bibnamefont {Yoshino}}, \bibinfo {author} {\bibfnamefont
  {K.}~\bibnamefont {Harii}}, \bibinfo {author} {\bibfnamefont
  {Y.}~\bibnamefont {Fujikawa}}, \bibinfo {author} {\bibfnamefont
  {M.}~\bibnamefont {Matsuo}}, \bibinfo {author} {\bibfnamefont
  {S.}~\bibnamefont {Maekawa}}, \ and\ \bibinfo {author} {\bibfnamefont
  {E.}~\bibnamefont {Saitoh}},\ }\href {https://dx.doi.org/10.1063/1.3587173}
  {\bibfield  {journal} {\bibinfo  {journal} {J. App. Phys.}\ }\textbf
  {\bibinfo {volume} {109}},\ \bibinfo {pages} {103913} (\bibinfo {year}
  {2010})}\BibitemShut {NoStop}%
\bibitem [{\citenamefont {Jungfleisch}\ \emph {et~al.}(2015)\citenamefont
  {Jungfleisch}, \citenamefont {Chumak}, \citenamefont {Kehlberger},
  \citenamefont {Lauer}, \citenamefont {Kim}, \citenamefont {Onbasli},
  \citenamefont {Ross}, \citenamefont {Kl{\"a}ui},\ and\ \citenamefont
  {Hillebrands}}]{jungfleischPRB2015}%
  \BibitemOpen
  \bibfield  {author} {\bibinfo {author} {\bibfnamefont {M.~B.}\ \bibnamefont
  {Jungfleisch}}, \bibinfo {author} {\bibfnamefont {A.~V.}\ \bibnamefont
  {Chumak}}, \bibinfo {author} {\bibfnamefont {A.}~\bibnamefont {Kehlberger}},
  \bibinfo {author} {\bibfnamefont {V.}~\bibnamefont {Lauer}}, \bibinfo
  {author} {\bibfnamefont {D.~H.}\ \bibnamefont {Kim}}, \bibinfo {author}
  {\bibfnamefont {M.}~\bibnamefont {Onbasli}}, \bibinfo {author} {\bibfnamefont
  {C.~A.}\ \bibnamefont {Ross}}, \bibinfo {author} {\bibfnamefont
  {M.}~\bibnamefont {Kl{\"a}ui}}, \ and\ \bibinfo {author} {\bibfnamefont
  {B.}~\bibnamefont {Hillebrands}},\ }\href
  {https://doi.org/10.1103/PhysRevB.91.134407} {\bibfield  {journal} {\bibinfo
  {journal} {Phys. Rev. B}\ }\textbf {\bibinfo {volume} {91}},\ \bibinfo
  {pages} {134407} (\bibinfo {year} {2015})}\BibitemShut {NoStop}%
\bibitem [{\citenamefont {Cornelissen}\ \emph {et~al.}(2010)\citenamefont
  {Cornelissen}, \citenamefont {Liu}, \citenamefont {Duine}, \citenamefont
  {Youssef},\ and\ \citenamefont {van Wees}}]{cornelissenNat2015}%
  \BibitemOpen
  \bibfield  {author} {\bibinfo {author} {\bibfnamefont {L.~J.}\ \bibnamefont
  {Cornelissen}}, \bibinfo {author} {\bibfnamefont {J.}~\bibnamefont {Liu}},
  \bibinfo {author} {\bibfnamefont {R.~A.}\ \bibnamefont {Duine}}, \bibinfo
  {author} {\bibfnamefont {J.~B.}\ \bibnamefont {Youssef}}, \ and\ \bibinfo
  {author} {\bibfnamefont {B.~J.}\ \bibnamefont {van Wees}},\ }\href
  {https://dx.doi.org/10.1038/nphys3465} {\bibfield  {journal} {\bibinfo
  {journal} {Nat. Phys.}\ }\textbf {\bibinfo {volume} {11}},\ \bibinfo {pages}
  {12} (\bibinfo {year} {2010})}\BibitemShut {NoStop}%
\bibitem [{\citenamefont {Avci}\ \emph {et~al.}(2016)\citenamefont {Avci},
  \citenamefont {Quindeau}, \citenamefont {Pai}, \citenamefont {Mann},
  \citenamefont {Caretta}, \citenamefont {Tang}, \citenamefont {Onbasli},
  \citenamefont {Ross},\ and\ \citenamefont {Beach}}]{avciNat2016}%
  \BibitemOpen
  \bibfield  {author} {\bibinfo {author} {\bibfnamefont {C.~O.}\ \bibnamefont
  {Avci}}, \bibinfo {author} {\bibfnamefont {A.}~\bibnamefont {Quindeau}},
  \bibinfo {author} {\bibfnamefont {C.}~\bibnamefont {Pai}}, \bibinfo {author}
  {\bibfnamefont {M.}~\bibnamefont {Mann}}, \bibinfo {author} {\bibfnamefont
  {L.}~\bibnamefont {Caretta}}, \bibinfo {author} {\bibfnamefont {A.~S.}\
  \bibnamefont {Tang}}, \bibinfo {author} {\bibfnamefont {M.~C.}\ \bibnamefont
  {Onbasli}}, \bibinfo {author} {\bibfnamefont {C.~A.}\ \bibnamefont {Ross}}, \
  and\ \bibinfo {author} {\bibfnamefont {G.~S.~D.}\ \bibnamefont {Beach}},\
  }\href {http://dx.doi.org/10.1038/nmat4812} {\bibfield  {journal} {\bibinfo
  {journal} {Nat. Mater.}\ }\textbf {\bibinfo {volume} {16}},\ \bibinfo {pages}
  {4812} (\bibinfo {year} {2016})}\BibitemShut {NoStop}%
\bibitem [{\citenamefont {Serga}\ \emph {et~al.}(2010)\citenamefont {Serga},
  \citenamefont {Chumak},\ and\ \citenamefont {Hillebrands}}]{sergaApp2010}%
  \BibitemOpen
  \bibfield  {author} {\bibinfo {author} {\bibfnamefont {A.~A.}\ \bibnamefont
  {Serga}}, \bibinfo {author} {\bibfnamefont {A.~V.}\ \bibnamefont {Chumak}}, \
  and\ \bibinfo {author} {\bibfnamefont {B.}~\bibnamefont {Hillebrands}},\
  }\href {http://stacks.iop.org/0022-3727/43/i=26/a=264002} {\bibfield
  {journal} {\bibinfo  {journal} {J. Phys. D.}\ }\textbf {\bibinfo {volume}
  {43}},\ \bibinfo {pages} {264002} (\bibinfo {year} {2010})}\BibitemShut
  {NoStop}%
\bibitem [{\citenamefont {Kirkpatrick}\ and\ \citenamefont
  {Belitz}(2003)}]{KirkpatrickPRB2003}%
  \BibitemOpen
  \bibfield  {author} {\bibinfo {author} {\bibfnamefont {T.~R.}\ \bibnamefont
  {Kirkpatrick}}\ and\ \bibinfo {author} {\bibfnamefont {D.}~\bibnamefont
  {Belitz}},\ }\href {https://doi.org/10.1103/PhysRevB.67.024515} {\bibfield
  {journal} {\bibinfo  {journal} {Phys. Rev. B}\ }\textbf {\bibinfo {volume}
  {67}},\ \bibinfo {pages} {024515} (\bibinfo {year} {2003})}\BibitemShut
  {NoStop}%
\bibitem [{\citenamefont {Karchev}(2003)}]{KarchevPRB2003}%
  \BibitemOpen
  \bibfield  {author} {\bibinfo {author} {\bibfnamefont {N.}~\bibnamefont
  {Karchev}},\ }\href {https://doi.org/10.1103/PhysRevB.67.054416} {\bibfield
  {journal} {\bibinfo  {journal} {Phys. Rev. B}\ }\textbf {\bibinfo {volume}
  {67}},\ \bibinfo {pages} {054416} (\bibinfo {year} {2003})}\BibitemShut
  {NoStop}%
\bibitem [{\citenamefont {Saxena}\ \emph {et~al.}(2000)\citenamefont {Saxena},
  \citenamefont {Agarwal}, \citenamefont {Ahilan}, \citenamefont {Grosche},
  \citenamefont {Haselwimmer}, \citenamefont {Steiner}, \citenamefont {Pugh},
  \citenamefont {Walker}, \citenamefont {Julian}, \citenamefont {Monthoux},
  \citenamefont {Lonzarich}, \citenamefont {Sheikin}, \citenamefont
  {Braithwaite},\ and\ \citenamefont {Flouquet}}]{SaxenaNat2000}%
  \BibitemOpen
  \bibfield  {author} {\bibinfo {author} {\bibfnamefont {S.~S.}\ \bibnamefont
  {Saxena}}, \bibinfo {author} {\bibfnamefont {P.}~\bibnamefont {Agarwal}},
  \bibinfo {author} {\bibfnamefont {K.}~\bibnamefont {Ahilan}}, \bibinfo
  {author} {\bibfnamefont {F.~M.}\ \bibnamefont {Grosche}}, \bibinfo {author}
  {\bibfnamefont {R.~K.~W.}\ \bibnamefont {Haselwimmer}}, \bibinfo {author}
  {\bibfnamefont {M.~J.}\ \bibnamefont {Steiner}}, \bibinfo {author}
  {\bibfnamefont {E.}~\bibnamefont {Pugh}}, \bibinfo {author} {\bibfnamefont
  {I.~R.}\ \bibnamefont {Walker}}, \bibinfo {author} {\bibfnamefont
  {S.}~\bibnamefont {Julian}}, \bibinfo {author} {\bibfnamefont
  {P.}~\bibnamefont {Monthoux}}, \bibinfo {author} {\bibfnamefont {G.~G.}\
  \bibnamefont {Lonzarich}}, \bibinfo {author} {\bibfnamefont {A.~H.~I.}\
  \bibnamefont {Sheikin}}, \bibinfo {author} {\bibfnamefont {D.}~\bibnamefont
  {Braithwaite}}, \ and\ \bibinfo {author} {\bibfnamefont {J.}~\bibnamefont
  {Flouquet}},\ }\href {http://dx.doi.org/10.1038/35020500} {\bibfield
  {journal} {\bibinfo  {journal} {Nature (London)}\ }\textbf {\bibinfo {volume}
  {406}},\ \bibinfo {pages} {6796} (\bibinfo {year} {2000})}\BibitemShut
  {NoStop}%
\bibitem [{\citenamefont {Aoki}\ \emph {et~al.}(2001)\citenamefont {Aoki},
  \citenamefont {Huxley}, \citenamefont {Ressouche}, \citenamefont
  {Braithwaite}, \citenamefont {Flouquet}, \citenamefont {Brison},
  \citenamefont {Lhotel},\ and\ \citenamefont {Paulsen}}]{aokiNat2001}%
  \BibitemOpen
  \bibfield  {author} {\bibinfo {author} {\bibfnamefont {D.}~\bibnamefont
  {Aoki}}, \bibinfo {author} {\bibfnamefont {A.}~\bibnamefont {Huxley}},
  \bibinfo {author} {\bibfnamefont {E.}~\bibnamefont {Ressouche}}, \bibinfo
  {author} {\bibfnamefont {D.}~\bibnamefont {Braithwaite}}, \bibinfo {author}
  {\bibfnamefont {J.}~\bibnamefont {Flouquet}}, \bibinfo {author}
  {\bibfnamefont {J.}~\bibnamefont {Brison}}, \bibinfo {author} {\bibfnamefont
  {E.}~\bibnamefont {Lhotel}}, \ and\ \bibinfo {author} {\bibfnamefont
  {C.}~\bibnamefont {Paulsen}},\ }\href {http://dx.doi.org/10.1038/35098048}
  {\bibfield  {journal} {\bibinfo  {journal} {Nature (London)}\ }\textbf
  {\bibinfo {volume} {413}},\ \bibinfo {pages} {6856} (\bibinfo {year}
  {2001})}\BibitemShut {NoStop}%
\bibitem [{\citenamefont {Pfleiderer}\ \emph {et~al.}(2001)\citenamefont
  {Pfleiderer}, \citenamefont {Uhlarz}, \citenamefont {Hayden}, \citenamefont
  {Vollmer}, \citenamefont {v.~L{\"o}hneysen}, \citenamefont {Bernhoeft},\ and\
  \citenamefont {Lonzarich}}]{pfleidererNat2001}%
  \BibitemOpen
  \bibfield  {author} {\bibinfo {author} {\bibfnamefont {C.}~\bibnamefont
  {Pfleiderer}}, \bibinfo {author} {\bibfnamefont {M.}~\bibnamefont {Uhlarz}},
  \bibinfo {author} {\bibfnamefont {S.~M.}\ \bibnamefont {Hayden}}, \bibinfo
  {author} {\bibfnamefont {R.}~\bibnamefont {Vollmer}}, \bibinfo {author}
  {\bibfnamefont {H.}~\bibnamefont {v.~L{\"o}hneysen}}, \bibinfo {author}
  {\bibfnamefont {N.~R.}\ \bibnamefont {Bernhoeft}}, \ and\ \bibinfo {author}
  {\bibfnamefont {G.~G.}\ \bibnamefont {Lonzarich}},\ }\href
  {http://dx.doi.org/10.1038/35083531} {\bibfield  {journal} {\bibinfo
  {journal} {Nature (London)}\ }\textbf {\bibinfo {volume} {412}},\ \bibinfo
  {pages} {6842} (\bibinfo {year} {2001})}\BibitemShut {NoStop}%
\bibitem [{\citenamefont {Fay}\ and\ \citenamefont {Appel}(1980)}]{FayPRB1980}%
  \BibitemOpen
  \bibfield  {author} {\bibinfo {author} {\bibfnamefont {D.}~\bibnamefont
  {Fay}}\ and\ \bibinfo {author} {\bibfnamefont {J.}~\bibnamefont {Appel}},\
  }\href {https://doi.org/10.1103/PhysRevB.22.3173} {\bibfield  {journal}
  {\bibinfo  {journal} {Phys. Rev. B}\ }\textbf {\bibinfo {volume} {22}},\
  \bibinfo {pages} {3173} (\bibinfo {year} {1980})}\BibitemShut {NoStop}%
\bibitem [{\citenamefont {Karchev}(2015)}]{KarchevEPL2015}%
  \BibitemOpen
  \bibfield  {author} {\bibinfo {author} {\bibfnamefont {N.}~\bibnamefont
  {Karchev}},\ }\href {http://stacks.iop.org/0295-5075/110/i=2/a=27004}
  {\bibfield  {journal} {\bibinfo  {journal} {EPL (Europhysics Letters)}\
  }\textbf {\bibinfo {volume} {110}},\ \bibinfo {pages} {27004} (\bibinfo
  {year} {2015})}\BibitemShut {NoStop}%
\bibitem [{\citenamefont {Saito}\ \emph {et~al.}(2016)\citenamefont {Saito},
  \citenamefont {Nojima},\ and\ \citenamefont {Iwasa}}]{saitoNat2016}%
  \BibitemOpen
  \bibfield  {author} {\bibinfo {author} {\bibfnamefont {Y.}~\bibnamefont
  {Saito}}, \bibinfo {author} {\bibfnamefont {T.}~\bibnamefont {Nojima}}, \
  and\ \bibinfo {author} {\bibfnamefont {Y.}~\bibnamefont {Iwasa}},\ }\href
  {https://dx.doi.org/10.1038/natrevmats.2016.94} {\bibfield  {journal}
  {\bibinfo  {journal} {Nat. Rev. Mater.}\ }\textbf {\bibinfo {volume} {2}},\
  \bibinfo {pages} {16094} (\bibinfo {year} {2016})}\BibitemShut {NoStop}%
\bibitem [{\citenamefont {Gariglio}\ \emph {et~al.}(2015)\citenamefont
  {Gariglio}, \citenamefont {Gabay}, \citenamefont {Mannhart},\ and\
  \citenamefont {Triscone}}]{Gariglioetal2015}%
  \BibitemOpen
  \bibfield  {author} {\bibinfo {author} {\bibfnamefont {S.}~\bibnamefont
  {Gariglio}}, \bibinfo {author} {\bibfnamefont {M.}~\bibnamefont {Gabay}},
  \bibinfo {author} {\bibfnamefont {J.}~\bibnamefont {Mannhart}}, \ and\
  \bibinfo {author} {\bibfnamefont {J.-M.}\ \bibnamefont {Triscone}},\ }\href
  {http://dx.doi.org/10.1016/j.physc.2015.02.028} {\bibfield  {journal}
  {\bibinfo  {journal} {Physica C}\ }\textbf {\bibinfo {volume} {514}},\
  \bibinfo {pages} {189} (\bibinfo {year} {2015})}\BibitemShut {NoStop}%
\bibitem [{\citenamefont {Reyren}\ \emph {et~al.}(2007)\citenamefont {Reyren},
  \citenamefont {Thiel}, \citenamefont {Caviglia}, \citenamefont {Kourkoutis},
  \citenamefont {Hammerl}, \citenamefont {Richter}, \citenamefont {Schneider},
  \citenamefont {Kopp}, \citenamefont {R{\"u}etschi}, \citenamefont {Jaccard},
  \citenamefont {Gabay}, \citenamefont {Muller}, \citenamefont {Triscone},\
  and\ \citenamefont {Mannhart}}]{ReyrenetalScience2007}%
  \BibitemOpen
  \bibfield  {author} {\bibinfo {author} {\bibfnamefont {N.}~\bibnamefont
  {Reyren}}, \bibinfo {author} {\bibfnamefont {S.}~\bibnamefont {Thiel}},
  \bibinfo {author} {\bibfnamefont {A.~D.}\ \bibnamefont {Caviglia}}, \bibinfo
  {author} {\bibfnamefont {L.~F.}\ \bibnamefont {Kourkoutis}}, \bibinfo
  {author} {\bibfnamefont {G.}~\bibnamefont {Hammerl}}, \bibinfo {author}
  {\bibfnamefont {C.}~\bibnamefont {Richter}}, \bibinfo {author} {\bibfnamefont
  {C.~W.}\ \bibnamefont {Schneider}}, \bibinfo {author} {\bibfnamefont
  {T.}~\bibnamefont {Kopp}}, \bibinfo {author} {\bibfnamefont {A.-S.}\
  \bibnamefont {R{\"u}etschi}}, \bibinfo {author} {\bibfnamefont
  {D.}~\bibnamefont {Jaccard}}, \bibinfo {author} {\bibfnamefont
  {M.}~\bibnamefont {Gabay}}, \bibinfo {author} {\bibfnamefont {D.~A.}\
  \bibnamefont {Muller}}, \bibinfo {author} {\bibfnamefont {J.-M.}\
  \bibnamefont {Triscone}}, \ and\ \bibinfo {author} {\bibfnamefont
  {J.}~\bibnamefont {Mannhart}},\ }\href
  {https://doi.org/10.1126/science.1146006} {\bibfield  {journal} {\bibinfo
  {journal} {Science}\ }\textbf {\bibinfo {volume} {317}},\ \bibinfo {pages}
  {1196} (\bibinfo {year} {2007})}\BibitemShut {NoStop}%
\bibitem [{\citenamefont {Wang}\ \emph {et~al.}(2012)\citenamefont {Wang},
  \citenamefont {Li}, \citenamefont {Zhang}, \citenamefont {Zhang},
  \citenamefont {Zhang}, \citenamefont {Li}, \citenamefont {Ding},
  \citenamefont {Ou}, \citenamefont {Deng}, \citenamefont {Chang},
  \citenamefont {Wen}, \citenamefont {Song}, \citenamefont {He}, \citenamefont
  {Jia}, \citenamefont {Ji}, \citenamefont {Wang}, \citenamefont {Wang},
  \citenamefont {Chen}, \citenamefont {Ma},\ and\ \citenamefont
  {Xue}}]{Wangetal2012}%
  \BibitemOpen
  \bibfield  {author} {\bibinfo {author} {\bibfnamefont {Q.~Y.}\ \bibnamefont
  {Wang}}, \bibinfo {author} {\bibfnamefont {Z.}~\bibnamefont {Li}}, \bibinfo
  {author} {\bibfnamefont {W.~H.}\ \bibnamefont {Zhang}}, \bibinfo {author}
  {\bibfnamefont {Z.~C.}\ \bibnamefont {Zhang}}, \bibinfo {author}
  {\bibfnamefont {J.~S.}\ \bibnamefont {Zhang}}, \bibinfo {author}
  {\bibfnamefont {W.}~\bibnamefont {Li}}, \bibinfo {author} {\bibfnamefont
  {H.}~\bibnamefont {Ding}}, \bibinfo {author} {\bibfnamefont {Y.~B.}\
  \bibnamefont {Ou}}, \bibinfo {author} {\bibfnamefont {P.}~\bibnamefont
  {Deng}}, \bibinfo {author} {\bibfnamefont {K.}~\bibnamefont {Chang}},
  \bibinfo {author} {\bibfnamefont {J.}~\bibnamefont {Wen}}, \bibinfo {author}
  {\bibfnamefont {C.~L.}\ \bibnamefont {Song}}, \bibinfo {author}
  {\bibfnamefont {K.}~\bibnamefont {He}}, \bibinfo {author} {\bibfnamefont
  {J.~F.}\ \bibnamefont {Jia}}, \bibinfo {author} {\bibfnamefont {S.~H.}\
  \bibnamefont {Ji}}, \bibinfo {author} {\bibfnamefont {Y.~Y.}\ \bibnamefont
  {Wang}}, \bibinfo {author} {\bibfnamefont {L.~L.}\ \bibnamefont {Wang}},
  \bibinfo {author} {\bibfnamefont {X.}~\bibnamefont {Chen}}, \bibinfo {author}
  {\bibfnamefont {X.~C.}\ \bibnamefont {Ma}}, \ and\ \bibinfo {author}
  {\bibfnamefont {Q.~K.}\ \bibnamefont {Xue}},\ }\href
  {http://doi.org/10.1088/0256-307X/29/3/037402} {\bibfield  {journal}
  {\bibinfo  {journal} {Chin.~Phys.~Lett.}\ }\textbf {\bibinfo {volume} {29}},\
  \bibinfo {pages} {037402} (\bibinfo {year} {2012})}\BibitemShut {NoStop}%
\bibitem [{\citenamefont {Boschker}\ \emph {et~al.}(2015)\citenamefont
  {Boschker}, \citenamefont {Richter}, \citenamefont {Fillis-Tsirakis},
  \citenamefont {Schneider},\ and\ \citenamefont
  {Mannhart}}]{Boschkeretal2015}%
  \BibitemOpen
  \bibfield  {author} {\bibinfo {author} {\bibfnamefont {H.}~\bibnamefont
  {Boschker}}, \bibinfo {author} {\bibfnamefont {C.}~\bibnamefont {Richter}},
  \bibinfo {author} {\bibfnamefont {E.}~\bibnamefont {Fillis-Tsirakis}},
  \bibinfo {author} {\bibfnamefont {C.~W.}\ \bibnamefont {Schneider}}, \ and\
  \bibinfo {author} {\bibfnamefont {J.}~\bibnamefont {Mannhart}},\ }\href
  {https://doi.org/10.1038/srep12309} {\bibfield  {journal} {\bibinfo
  {journal} {Sci.~Rep.}\ }\textbf {\bibinfo {volume} {5}},\ \bibinfo {pages}
  {12309} (\bibinfo {year} {2015})}\BibitemShut {NoStop}%
\bibitem [{\citenamefont {Klimin}\ \emph {et~al.}(2014)\citenamefont {Klimin},
  \citenamefont {Tempere}, \citenamefont {Devreese},\ and\ \citenamefont
  {van~der Marel}}]{Kliminetal2014}%
  \BibitemOpen
  \bibfield  {author} {\bibinfo {author} {\bibfnamefont {S.~N.}\ \bibnamefont
  {Klimin}}, \bibinfo {author} {\bibfnamefont {J.}~\bibnamefont {Tempere}},
  \bibinfo {author} {\bibfnamefont {J.~T.}\ \bibnamefont {Devreese}}, \ and\
  \bibinfo {author} {\bibfnamefont {D.}~\bibnamefont {van~der Marel}},\ }\href
  {http://doi.org/10.1103/PhysRevB.89.184514} {\bibfield  {journal} {\bibinfo
  {journal} {Phys.~Rev.~B}\ }\textbf {\bibinfo {volume} {89}},\ \bibinfo
  {pages} {184514} (\bibinfo {year} {2014})}\BibitemShut {NoStop}%
\bibitem [{\citenamefont {Li}\ \emph {et~al.}(2014)\citenamefont {Li},
  \citenamefont {Xing}, \citenamefont {Huang},\ and\ \citenamefont
  {Xing}}]{Lieetal2014}%
  \BibitemOpen
  \bibfield  {author} {\bibinfo {author} {\bibfnamefont {B.}~\bibnamefont
  {Li}}, \bibinfo {author} {\bibfnamefont {Z.~W.}\ \bibnamefont {Xing}},
  \bibinfo {author} {\bibfnamefont {G.~Q.}\ \bibnamefont {Huang}}, \ and\
  \bibinfo {author} {\bibfnamefont {D.~Y.}\ \bibnamefont {Xing}},\ }\href
  {http://doi.org/10.1063/1.4876750} {\bibfield  {journal} {\bibinfo  {journal}
  {J.~Appl.~Phys.}\ }\textbf {\bibinfo {volume} {115}},\ \bibinfo {pages}
  {193907} (\bibinfo {year} {2014})}\BibitemShut {NoStop}%
\bibitem [{\citenamefont {Gong}\ \emph {et~al.}(2015)\citenamefont {Gong},
  \citenamefont {Zhou}, \citenamefont {Xu}, \citenamefont {Yue}, \citenamefont
  {Zhu}, \citenamefont {Jin}, \citenamefont {Tian}, \citenamefont {Zhao},\ and\
  \citenamefont {Chen}}]{GongCPL2015}%
  \BibitemOpen
  \bibfield  {author} {\bibinfo {author} {\bibfnamefont {X.-X.}\ \bibnamefont
  {Gong}}, \bibinfo {author} {\bibfnamefont {H.-X.}\ \bibnamefont {Zhou}},
  \bibinfo {author} {\bibfnamefont {P.-C.}\ \bibnamefont {Xu}}, \bibinfo
  {author} {\bibfnamefont {D.}~\bibnamefont {Yue}}, \bibinfo {author}
  {\bibfnamefont {K.}~\bibnamefont {Zhu}}, \bibinfo {author} {\bibfnamefont
  {X.-F.}\ \bibnamefont {Jin}}, \bibinfo {author} {\bibfnamefont
  {H.}~\bibnamefont {Tian}}, \bibinfo {author} {\bibfnamefont {G.-J.}\
  \bibnamefont {Zhao}}, \ and\ \bibinfo {author} {\bibfnamefont {T.-Y.}\
  \bibnamefont {Chen}},\ }\href
  {http://stacks.iop.org/0256-307X/32/i=6/a=067402} {\bibfield  {journal}
  {\bibinfo  {journal} {Chinese Physics Letters}\ }\textbf {\bibinfo {volume}
  {32}},\ \bibinfo {pages} {067402} (\bibinfo {year} {2015})}\BibitemShut
  {NoStop}%
\bibitem [{\citenamefont {Stephanos}\ \emph {et~al.}(2011)\citenamefont
  {Stephanos}, \citenamefont {Kopp}, \citenamefont {Mannhart},\ and\
  \citenamefont {Hirschfeld}}]{StephanosetalPRB2011}%
  \BibitemOpen
  \bibfield  {author} {\bibinfo {author} {\bibfnamefont {C.}~\bibnamefont
  {Stephanos}}, \bibinfo {author} {\bibfnamefont {T.}~\bibnamefont {Kopp}},
  \bibinfo {author} {\bibfnamefont {J.}~\bibnamefont {Mannhart}}, \ and\
  \bibinfo {author} {\bibfnamefont {P.~J.}\ \bibnamefont {Hirschfeld}},\ }\href
  {https://doi.org/10.1103/PhysRevB.84.100510} {\bibfield  {journal} {\bibinfo
  {journal} {Phys.~Rev.~B}\ }\textbf {\bibinfo {volume} {84}},\ \bibinfo
  {pages} {100510(R)} (\bibinfo {year} {2011})}\BibitemShut {NoStop}%
\bibitem [{\citenamefont {Allender}\ \emph {et~al.}(1973)\citenamefont
  {Allender}, \citenamefont {Bray},\ and\ \citenamefont
  {Bardeen}}]{Allenderetal1973}%
  \BibitemOpen
  \bibfield  {author} {\bibinfo {author} {\bibfnamefont {D.}~\bibnamefont
  {Allender}}, \bibinfo {author} {\bibfnamefont {J.}~\bibnamefont {Bray}}, \
  and\ \bibinfo {author} {\bibfnamefont {J.}~\bibnamefont {Bardeen}},\ }\href
  {https://doi.org/10.1103/PhysRevB.7.1020} {\bibfield  {journal} {\bibinfo
  {journal} {Phys.~Rev.~B}\ }\textbf {\bibinfo {volume} {7}},\ \bibinfo {pages}
  {1020} (\bibinfo {year} {1973})}\BibitemShut {NoStop}%
\bibitem [{\citenamefont {Koerting}\ \emph {et~al.}(2005)\citenamefont
  {Koerting}, \citenamefont {Yuan}, \citenamefont {Hirschfeld}, \citenamefont
  {Kopp},\ and\ \citenamefont {Mannhart}}]{KoertingetalPRB2005}%
  \BibitemOpen
  \bibfield  {author} {\bibinfo {author} {\bibfnamefont {V.}~\bibnamefont
  {Koerting}}, \bibinfo {author} {\bibfnamefont {Q.}~\bibnamefont {Yuan}},
  \bibinfo {author} {\bibfnamefont {P.~J.}\ \bibnamefont {Hirschfeld}},
  \bibinfo {author} {\bibfnamefont {T.}~\bibnamefont {Kopp}}, \ and\ \bibinfo
  {author} {\bibfnamefont {J.}~\bibnamefont {Mannhart}},\ }\href
  {https://doi.org/10.1103/PhysRevB.71.104510} {\bibfield  {journal} {\bibinfo
  {journal} {Phys.~Rev.~B}\ }\textbf {\bibinfo {volume} {71}},\ \bibinfo
  {pages} {104510} (\bibinfo {year} {2005})}\BibitemShut {NoStop}%
\bibitem [{\citenamefont {Gong}\ \emph {et~al.}(2017)\citenamefont {Gong},
  \citenamefont {Kargarian}, \citenamefont {Stern}, \citenamefont {Yue},
  \citenamefont {Zhou}, \citenamefont {Jin}, \citenamefont {Galitski},
  \citenamefont {Yakovenko},\ and\ \citenamefont {Xia}}]{GongeSCIENCE2017}%
  \BibitemOpen
  \bibfield  {author} {\bibinfo {author} {\bibfnamefont {X.}~\bibnamefont
  {Gong}}, \bibinfo {author} {\bibfnamefont {M.}~\bibnamefont {Kargarian}},
  \bibinfo {author} {\bibfnamefont {A.}~\bibnamefont {Stern}}, \bibinfo
  {author} {\bibfnamefont {D.}~\bibnamefont {Yue}}, \bibinfo {author}
  {\bibfnamefont {H.}~\bibnamefont {Zhou}}, \bibinfo {author} {\bibfnamefont
  {X.}~\bibnamefont {Jin}}, \bibinfo {author} {\bibfnamefont {V.~M.}\
  \bibnamefont {Galitski}}, \bibinfo {author} {\bibfnamefont {V.~M.}\
  \bibnamefont {Yakovenko}}, \ and\ \bibinfo {author} {\bibfnamefont
  {J.}~\bibnamefont {Xia}},\ }\href {\doibase 10.1126/sciadv.1602579}
  {\bibfield  {journal} {\bibinfo  {journal} {Science Advances}\ }\textbf
  {\bibinfo {volume} {3}} (\bibinfo {year} {2017}),\
  10.1126/sciadv.1602579}\BibitemShut {NoStop}%
\bibitem [{\citenamefont {Kargarian}\ \emph {et~al.}(2016)\citenamefont
  {Kargarian}, \citenamefont {Efimkin},\ and\ \citenamefont
  {Galitski}}]{KagarianPRL2016}%
  \BibitemOpen
  \bibfield  {author} {\bibinfo {author} {\bibfnamefont {M.}~\bibnamefont
  {Kargarian}}, \bibinfo {author} {\bibfnamefont {D.~K.}\ \bibnamefont
  {Efimkin}}, \ and\ \bibinfo {author} {\bibfnamefont {V.}~\bibnamefont
  {Galitski}},\ }\href {\doibase 10.1103/PhysRevLett.117.076806} {\bibfield
  {journal} {\bibinfo  {journal} {Phys. Rev. Lett.}\ }\textbf {\bibinfo
  {volume} {117}},\ \bibinfo {pages} {076806} (\bibinfo {year}
  {2016})}\BibitemShut {NoStop}%
\bibitem [{Note1()}]{Note1}%
  \BibitemOpen
  \bibinfo {note} {For the integration, we use the Cauchy principle value and
  the fact that $\Delta x=|x-x'|\ll 1$,
  \begin {equation*} \begin {split}
  V(\Delta x) = & - \protect \frac {\protect \sqrt {2}}{\pi }\DOTSI \intop
  \ilimits@ _0^{2\pi }d\protect \mathaccentV {tilde}07E\varphi \protect
  \tmspace +\thinmuskip {.1667em} \protect \frac {(1+\protect \qopname \relax
  o{cos}\protect \mathaccentV {tilde}07E\varphi )\protect \qopname \relax
  o{cos}\protect \mathaccentV {tilde}07E\varphi } {(1+\protect \qopname \relax
  o{cos}\protect \mathaccentV {tilde}07E\varphi )^2-\Delta x^2}\\ = & \protect
  \frac {1+|\Delta x|}{\protect \sqrt {\Delta x^2/2+\Delta x}} - 2\protect
  \sqrt {2} \approx \protect \frac {1}{\protect \sqrt {\Delta x}}-2\protect
  \sqrt {2} \protect \tmspace +\thinmuskip {.1667em} . 
  \end {split}
  \end{equation*} \protect \trick .}\BibitemShut {Stop}%
\bibitem [{Note2()}]{Note2}%
  \BibitemOpen
  \bibinfo {note} {To eliminate the singularity in $V(x{-}x')$ at $x'=x$ for
  the numerical integration, we replace $V(x{-}x')$ in Eq.~(\ref {eq:gap_simp})
  with $\DOTSI \intop \ilimits@ _0^x d\protect \mathaccentV {tilde}07E{x}
  V(\protect \mathaccentV {tilde}07E{x}{-}x')$ and $f(x)$ on the left-hand side
  of Eq.~(\ref {eq:gap_simp}) with $F(x) = \DOTSI \intop \ilimits@ _0^x
  d\protect \mathaccentV {tilde}07E{x} f(\protect \mathaccentV {tilde}07E{x})$.
  In each iteration, we numerically evaluate the integral over $x'$ in the
  resulting equation and obtain $f(x)$ by numerically differentiating
  $F$.}\BibitemShut {Stop}%
\bibitem [{Note3()}]{Note3}%
  \BibitemOpen
  \bibinfo {note} {In the same way as for the iteration of Eq.~(\ref
  {eq:gap_simp}), we eliminate singularities from the integral over $x'$ in
  Eq.~(\ref {eq:gap_int}). Hence, we replace the factor
  $U(\varphi ',\varphi,x',x) = 1/[\epsilon ^2(\varphi ',\varphi )-(x{-}x')^2]$
  in the integrand with
  $\DOTSI \intop \ilimits@ _{0}^{x} d\protect \mathaccentV {tilde}07E{X}
  \DOTSI \intop \ilimits@ _{0}^{\protect \mathaccentV {tilde}07E{X}} d\protect
  \mathaccentV {tilde}07E{x} U(\varphi ',\varphi ,x',\protect \mathaccentV
  {tilde}07E{x})$
  and $\delta (x,\varphi )$ with
  $D(x,\varphi ) = \DOTSI \intop
  \ilimits@ _{0}^{x} d\protect \mathaccentV {tilde}07E{X} \DOTSI \intop
  \ilimits@ _{0}^{\protect \mathaccentV {tilde}07E{X}} d\protect \mathaccentV
  {tilde}07E{x} \delta (\protect \mathaccentV {tilde}07E{x},\varphi )$.
  In each iteration, we find $D$ by numerically integrating over $x'$ and then find
  $\delta $ by numerically differentiating $D$ twice.}\BibitemShut {Stop}%
\bibitem [{\citenamefont {Klingler}\ \emph {et~al.}(2015)\citenamefont
  {Klingler}, \citenamefont {Chumak}, \citenamefont {Mewes}, \citenamefont
  {Khodadadi}, \citenamefont {Mewes}, \citenamefont {Dubs}, \citenamefont
  {Surzhenko}, \citenamefont {Hillebrands},\ and\ \citenamefont
  {Conca}}]{klinglerApp2015}%
  \BibitemOpen
  \bibfield  {author} {\bibinfo {author} {\bibfnamefont {S.}~\bibnamefont
  {Klingler}}, \bibinfo {author} {\bibfnamefont {A.~V.}\ \bibnamefont
  {Chumak}}, \bibinfo {author} {\bibfnamefont {T.}~\bibnamefont {Mewes}},
  \bibinfo {author} {\bibfnamefont {B.}~\bibnamefont {Khodadadi}}, \bibinfo
  {author} {\bibfnamefont {C.}~\bibnamefont {Mewes}}, \bibinfo {author}
  {\bibfnamefont {C.}~\bibnamefont {Dubs}}, \bibinfo {author} {\bibfnamefont
  {O.}~\bibnamefont {Surzhenko}}, \bibinfo {author} {\bibfnamefont
  {B.}~\bibnamefont {Hillebrands}}, \ and\ \bibinfo {author} {\bibfnamefont
  {A.}~\bibnamefont {Conca}},\ }\href
  {http://stacks.iop.org/0022-3727/48/i=1/a=015001} {\bibfield  {journal}
  {\bibinfo  {journal} {J. Phys. D.}\ }\textbf {\bibinfo {volume} {48}},\
  \bibinfo {pages} {015001} (\bibinfo {year} {2015})}\BibitemShut {NoStop}%
\bibitem [{\citenamefont {Ashcroft}\ and\ \citenamefont
  {Mermin}(1976)}]{ashcroft1976}%
  \BibitemOpen
  \bibfield  {author} {\bibinfo {author} {\bibfnamefont {N.~W.}\ \bibnamefont
  {Ashcroft}}\ and\ \bibinfo {author} {\bibfnamefont {N.~D.}\ \bibnamefont
  {Mermin}},\ }\href {http://cds.cern.ch/record/102652} {\bibinfo {title} {\it Solid State Physics}}\ (\bibinfo
  {publisher} {Thomson},\ \bibinfo {year} {1976})\BibitemShut {NoStop}%
\bibitem [{\citenamefont {Bender}\ \emph {et~al.}(2012)\citenamefont {Bender},
  \citenamefont {Duine},\ and\ \citenamefont {Tserkovnyak}}]{benderPRL2012}%
  \BibitemOpen
  \bibfield  {author} {\bibinfo {author} {\bibfnamefont {S.~A.}\ \bibnamefont
  {Bender}}, \bibinfo {author} {\bibfnamefont {R.~A.}\ \bibnamefont {Duine}}, \
  and\ \bibinfo {author} {\bibfnamefont {Y.}~\bibnamefont {Tserkovnyak}},\
  }\href {https://doi.org/10.1103/PhysRevLett.108.246601} {\bibfield  {journal}
  {\bibinfo  {journal} {Phys. Rev. Lett.}\ }\textbf {\bibinfo {volume} {108}},\
  \bibinfo {pages} {246601} (\bibinfo {year} {2012})}\BibitemShut {NoStop}%
\bibitem [{\citenamefont {Heinrich}\ \emph {et~al.}(2011)\citenamefont
  {Heinrich}, \citenamefont {Burrowes}, \citenamefont {Montoya}, \citenamefont
  {Kardasz}, \citenamefont {Girt}, \citenamefont {Song}, \citenamefont {Sun},\
  and\ \citenamefont {Wu}}]{heinrichPRL2011}%
  \BibitemOpen
  \bibfield  {author} {\bibinfo {author} {\bibfnamefont {B.}~\bibnamefont
  {Heinrich}}, \bibinfo {author} {\bibfnamefont {C.}~\bibnamefont {Burrowes}},
  \bibinfo {author} {\bibfnamefont {E.}~\bibnamefont {Montoya}}, \bibinfo
  {author} {\bibfnamefont {B.}~\bibnamefont {Kardasz}}, \bibinfo {author}
  {\bibfnamefont {E.}~\bibnamefont {Girt}}, \bibinfo {author} {\bibfnamefont
  {Y.~Y.}\ \bibnamefont {Song}}, \bibinfo {author} {\bibfnamefont
  {Y.}~\bibnamefont {Sun}}, \ and\ \bibinfo {author} {\bibfnamefont
  {M.}~\bibnamefont {Wu}},\ }\href
  {https://doi.org/10.1103/PhysRevLett.107.066604} {\bibfield  {journal}
  {\bibinfo  {journal} {Phys. Rev. Lett.}\ }\textbf {\bibinfo {volume} {107}},\
  \bibinfo {pages} {066604} (\bibinfo {year} {2011})}\BibitemShut {NoStop}%
\bibitem [{\citenamefont {Burrowes}\ \emph {et~al.}(2012)\citenamefont
  {Burrowes}, \citenamefont {Heinrich}, \citenamefont {Kardasz}, \citenamefont
  {Montoya}, \citenamefont {Girt}, \citenamefont {Sun}, \citenamefont {Song},\
  and\ \citenamefont {Wu}}]{BurrowesApp2012}%
  \BibitemOpen
  \bibfield  {author} {\bibinfo {author} {\bibfnamefont {C.}~\bibnamefont
  {Burrowes}}, \bibinfo {author} {\bibfnamefont {B.}~\bibnamefont {Heinrich}},
  \bibinfo {author} {\bibfnamefont {B.}~\bibnamefont {Kardasz}}, \bibinfo
  {author} {\bibfnamefont {E.~.}\ \bibnamefont {Montoya}}, \bibinfo {author}
  {\bibfnamefont {E.}~\bibnamefont {Girt}}, \bibinfo {author} {\bibfnamefont
  {Y.}~\bibnamefont {Sun}}, \bibinfo {author} {\bibfnamefont {Y.}~\bibnamefont
  {Song}}, \ and\ \bibinfo {author} {\bibfnamefont {M.}~\bibnamefont {Wu}},\
  }\href {http://dx.doi.org/10.1063/1.3690918} {\bibfield  {journal} {\bibinfo
  {journal} {App. Phys. Lett.}\ }\textbf {\bibinfo {volume} {100}},\ \bibinfo
  {pages} {092403} (\bibinfo {year} {2012})}\BibitemShut {NoStop}%
\bibitem [{\citenamefont {Haertinger}\ \emph {et~al.}(2015)\citenamefont
  {Haertinger}, \citenamefont {Back}, \citenamefont {Lotze}, \citenamefont
  {Weiler}, \citenamefont {Gepr{\"a}gs}, \citenamefont {Huebl}, \citenamefont
  {Goennenwein},\ and\ \citenamefont {Woltersdorf}}]{haertingerPRB2015}%
  \BibitemOpen
  \bibfield  {author} {\bibinfo {author} {\bibfnamefont {M.}~\bibnamefont
  {Haertinger}}, \bibinfo {author} {\bibfnamefont {C.~H.}\ \bibnamefont
  {Back}}, \bibinfo {author} {\bibfnamefont {J.}~\bibnamefont {Lotze}},
  \bibinfo {author} {\bibfnamefont {M.}~\bibnamefont {Weiler}}, \bibinfo
  {author} {\bibfnamefont {S.}~\bibnamefont {Gepr{\"a}gs}}, \bibinfo {author}
  {\bibfnamefont {H.}~\bibnamefont {Huebl}}, \bibinfo {author} {\bibfnamefont
  {S.~T.~B.}\ \bibnamefont {Goennenwein}}, \ and\ \bibinfo {author}
  {\bibfnamefont {G.}~\bibnamefont {Woltersdorf}},\ }\href
  {https://doi.org/10.1103/PhysRevB.92.054437} {\bibfield  {journal} {\bibinfo
  {journal} {Phys. Rev. B}\ }\textbf {\bibinfo {volume} {92}},\ \bibinfo
  {pages} {054437} (\bibinfo {year} {2015})}\BibitemShut {NoStop}%
\bibitem [{\citenamefont {Mauger}\ and\ \citenamefont
  {Godart}(1986)}]{Mauger_Godart_1986}%
  \BibitemOpen
  \bibfield  {author} {\bibinfo {author} {\bibfnamefont {A.}~\bibnamefont
  {Mauger}}\ and\ \bibinfo {author} {\bibfnamefont {C.}~\bibnamefont
  {Godart}},\ }\href {https://doi.org/10.1016/0370-1573(86)90139-0} {\bibfield
  {journal} {\bibinfo  {journal} {Physics Reports}\ }\textbf {\bibinfo {volume}
  {141}},\ \bibinfo {pages} {51} (\bibinfo {year} {1986})}\BibitemShut
  {NoStop}%
\bibitem [{Note4()}]{Note4}%
  \BibitemOpen
  \bibinfo {note} {Note that the strength of the exchange coupling is defined
  differently in Refs.~\cite {Tkaczyk_thesis,Roesler_et_al_1994}; the values
  reported there must be divided by 2 to obtain the value of $J_I$ as it is
  defined in this Letter.}\BibitemShut {Stop}%
\end{thebibliography}
\end{document}